\documentclass[%
 aip,
 amsmath,amssymb,
 reprint,%
]{revtex4-1}

\usepackage{graphicx}
\usepackage{dcolumn}
\usepackage{bm}

\usepackage{filecontents}

\usepackage{natbib}

\begin{document}

\preprint{AIP/123-QED}

\title{Effects of optical attenuation, heat diffusion and acoustic coherence in photoacoustic signals produced by nanoparticles}

\author{J.~E. Alba--Rosales}
\affiliation{%
Divisi\'on de Ciencias e Ingenier\'{\i}as, Universidad de Guanajuato. Le\'on, Gto., M\'exico.
}%

\author{G. Ramos--Ortiz}
\affiliation{Centro de Investigaciones en \'Optica.  Le\'{o}n, Gto., M\'exico.}

\author{L.~F. Escamilla--Herrera}%
\affiliation{%
Instituto de Ciencias Nucleares, UNAM, CDMX, M\'exico.
}%

\author{B. Reyes--Ram\'{\i}rez}
\affiliation{Centro de Investigaciones en \'Optica.  Le\'{o}n, Gto., M\'exico.}

\author{L. Polo--Parada}
\email{poloparadal@missouri.edu}
\affiliation{Dalton Cardiovascular Research Center, Department of Medical Pharmacology and Physiology, University of Missouri-Columbia, Columbia, MO, USA.}

\author{G. Guti\'errez--Ju\'arez}
\email{ggutj@fisica.ugto.mx}
\affiliation{%
Divisi\'on de Ciencias e Ingenier\'{\i}as, Universidad de Guanajuato. Le\'on, Gto., M\'exico.
}%

\date{\today}

\begin{abstract}
Behavior of the photoacoustic signal produced by nanoparticles as a function of their concentration was studied in detail. As the concentration of nanoparticles is increased in a sample, the peak-to-peak photoacoustic amplitude increases linearly up to a certain value, after which an asymptotic saturated behavior is observed. To elucidate the mechanisms responsible for these observations, we evaluate the effects of nanoparticles concentration, the optical attenuation and the effects of heat propagation from nano-sources to their surroundings. We found that the saturation effect of the photoacoustic signal as a function the concentration of nanoparticles is explained by a combination of two different mechanisms. As has been suggested previously, but not modeled correctly, the most important mechanism is attributed to optical attenuation. The second mechanism is due to an interference destructive process attributed to the superimposition of the photoacoustic amplitudes generated for each nanoparticle, this explanation is reinforced through our experimental and simulations results; based on this, it is found that the linear behavior of the photoacoustic amplitude could be restricted to optical densities $\le0.5$.
\end{abstract}

                              
\maketitle

In recent years, pulsed laser-induced ultrasound (US), better known as the Photoacoustic (PA) effect, has had a major resurgence because its wide range of applications, mainly in the biological and medical areas~\cite{Chen:2015}, for instance, PA imaging~\cite{Chaigne:2016} and as monitoring method in thermo-therapy of cancer~\cite{Huang:2013}. Further the analogies between optical and acoustic phenomena, led to advancements in confocal PA microscopy~\cite{Wang:2014}, creations of new methodologies to detect US~\cite{Chao:2007} and generation of new materials to achieve thermal and/or acoustic contrasts~\cite{Gao:2016}. PA effect is produced by the absorption of CW modulated pulsed optical radiation by a medium. This absorption raises non-radiative decays that increase the temperature and causes mechanical waves typically in the range of US. The major advantages for PA techniques are their sensibility to distinguish different optical contrast and the US penetration in the tissue~\cite{Wang:2009}. 

Nowadays metallic nanoparticles (NPs) play an important role as enhancers of the PA signal, the design and application of these materials is subject to the type of applications desired, as well as to the available laser source.~\cite{Chen:2015,Zhong:2015,Bayer:2013}. 

Previous reports have shown that the PA amplitude is not always proportional to materials concentration; for example, the first reports of PA saturation are described in the work of J. W. Pin \cite{WPLin:1979} and J. Wang \cite{WangJ:2010}; in the later, authors employ a continuous absorption model to explain the saturation in red ink, but such model depends of multiple parameters (without physical meaning) adjustment to correctly fit the experimental data. Notwithstanding this deficiency, this reference has been the explanation for the PA saturation obtained in dyes, pigments and tissue. On the other hand, despite the different physical properties exhibit by NPs, a similar saturation effect is observed. In several cases a linear dependence between amplitude and concentration has been obtained; for example, gold (Au) nano-vesicles~\cite{Huang:2013}, Au nano-rods~\cite{Bayer:2013}, Au bio-conjugated nano-spheres~\cite{Mallidi:2009,Chamberland:2008}, silica-coated Au nano-rods~\cite{Jokerst:2012}, Au nano-carbon-tubes and $\text{Fe}_2\text{O}_3$ nano-spheres~\cite{Galanzha:2009}; while, in many other examples a saturated behavior is observed; Au nano-spheres~\cite{Khosroshahi:2015}, Au nano-cages~\cite{Song:2009} and Au nano-beacons~\cite{Pan:2010}. The origin of this discrepancy has been never analyzed in detail and there are few reports where the PA response from a discretized system in micro scale was obtained \cite{Karpiouk:2008,Saha:2011,Solano:2012}. 

Also, the PA signal shape at low concentrations has been observed that is symmetric whereas at high concentrations became asymmetric; this behavior has been extensively reported in the literature~\cite{Wang:2014,Solano:2012,Karpiouk:2008,Chamberland:2008,Sigrist:1978,Dixon:2015,Feis:2014,Pan:2012,Puxiang:2015,Emelianov:2009,Zhang:2015}. Therefore, the goal of this study is to explain the causes of these concentration-dependent effects. Herein it is proposed that the key characteristics of the nonlinear saturated behavior in the peak-to-peak (P-P) PA amplitude and its asymmetry at high  concentrations are explained by means of a photo-thermo-acoustic model \cite{YTian:2015}. Sigrist~\cite{Sigrist:1978} previously reported a model that explained the PA signal produced by a continuous media (liquids); in this paper, the model is extended to a discrete case from continuous media to NPs. The NPs are represented by single point that absorbs the incident radiation instantaneously \cite{Chen:2012}. With this model a simulation code was developed taking in account the heat source size, the light attenuated by the NPs and the coherence of a single PA signal source \cite{YTian:2015,TLee:2016}. The light losses, by absorption and scattering, can be described by the well-known Beer-Lambert (BL) law, which corresponds to an exponential decay in the amount of light. We hypothesize that the asymmetric shape of the PA signals is due to the interference of the individual PA signals generated by the NPs. To verify this, it is theoretically estimated the spatial region where interference of two PA waves occurs. These results are of great importance in applications where the NPs are used as a PA enhancer, since they allow to identify a dynamic range for PA amplitude generation and an optimal contrast agent concentration at which the maximum PA signal contrast is possible.

For this, PA experiments were conducted with 5, 10 and 100 nm Au NPs (spheres,  nanoComposix) in 2 mM sodium citrate dihydrate aqueous solution (Detailed information about the samples characteristics can be found in supplementary information). The suspensions optical density (OD) \cite{HBWeber:2003,HBMobley:1995} were measured as a function of the NPs concentration for each aliquot by using an UV-VIS spectrometer (Lambda 900 UV/VIS/NIR, Pekin Elmer). The experimental setup for the detection of PA signals is shown in Figure \ref{Fig1}. The second harmonic from a Nd:YAG pulsed laser (Brilliant, Quantel) was employed to provide 532 nm light with a pulse duration of 10 ns and a repetition rate of 10 Hz. This beam was focused into the PA cell using a couple of lenses such that a large Rayleigh waits length was obtained to be approximately constant inside the cell (0.7 mm of diameter). The energy per pulse was set at 1 mJ ($\pm5\%$, SD). The laser beam was set perpendicular to the transducer 2 mm away. The laser energy was monitored using a thermopile (1917-R, Newport) coupled to a power meter (818P-030-19, Newport). PA signals were detected with a homemade transducer (20 MHz central response and a bandwidth of 20 MHz)~\cite{Reyes:2014} and displayed by a 200-MHz oscilloscope (TDS5104B, Tektronix, Wilsonville, OR) triggered by a photo-diode (DET10A; Thorlabs, Newton, NJ) with a 1-nanoseconds rise time. The signals were amplified with a gain of 25 dB via 500 MHz amplifier (ZFL-500LN-BNC+, Mini-Circuits). NPs suspensions were diluted from stock concentration ($100\%$) in steps of $10\%$ using a sodium citrate aqueous solution at 2 µM (6028, Karal). 
\begin{figure}[h!]
	\centering
	\includegraphics[width=0.48\textwidth]{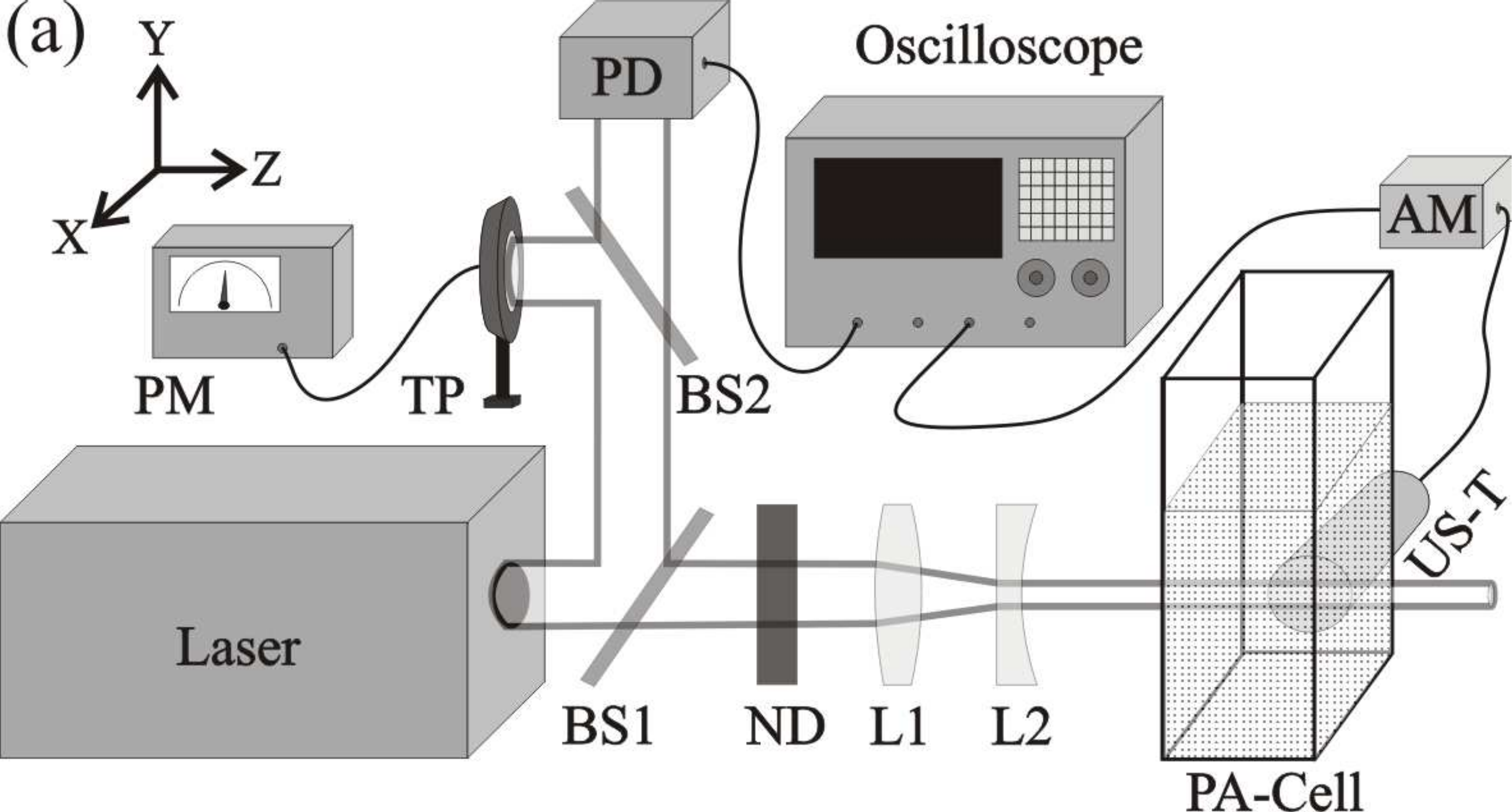}\label{Fig1}
	\caption{Experimental setup.PD: photo-diode, AM: amplifier, PM: power meter, TP: thermopile, ND: neutral density filters, BS1: beam spliter 10:90, BS2: beam spliter 10:90, L1: plane-convex lens with $f=+300$ mm, L2: plane-concave lens with $f=-50$ mm, US-T: ultra sound transducer. The distance between L1 and L2 is 250 mm.}
	\label{Fig1}
\end{figure}

Figure \ref{Fig2}(a) shows the PA signals generated by the NPs suspensions at stock concentration, these are $4 \times 10^{13}$ for $5$ nm, $5 \times 10^{12}$ for $10$ nm and $5 \times 10^{9}$ for $100$ nm. The signals generated by the NPs are observed at 1.3 $\mu$s. For lower NPs concentrations, the PA signals exhibited similar shape, but smaller amplitudes. Figures 2 (b) to 2 (d) show the respective normalized P-P PA amplitudes as a function of the NPs concentration for each NP size. 
\begin{figure}[h!]
	\centering
	\includegraphics[width=0.48\textwidth]{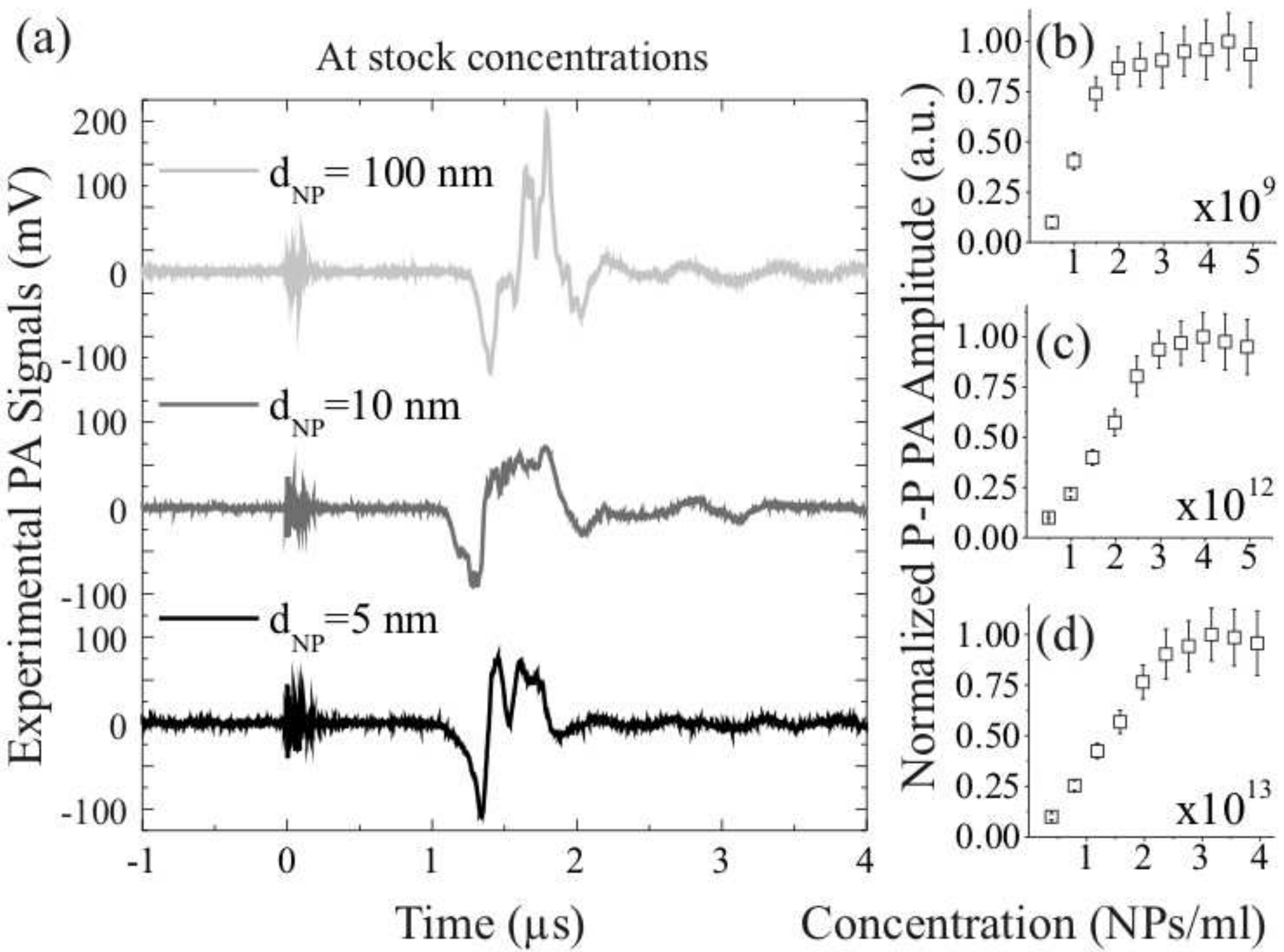}\label{Fig2}
	\caption{(a) Measured PA signals obtained from different NPs samples at stock concentration. (b)-(d) Normalized PA amplitude as a function of concentration for each NPs of 100 nm, 10 nm and 5 nm, respectively.}
    \label{Fig2}
\end{figure}

Here, it is assumed that the measured pressure $p(\mathbf{R},t)$ in $\mathbf{r}$ generated by a single nanoparticle (NP) at $\mathbf{r}'$ is described by  ~\cite{Chen:2012,Morse,Binbin:2013,Sigrist:1978}:

\begin{equation}\label{eq01}
p(\mathbf{R},t)=p_0\frac{ct-|\mathbf{R}|}{|\mathbf{R}|}\exp\Bigg[-\Bigg(\frac{ct-|\mathbf{R}|}{d_0/2}\Bigg)^2\Bigg];
\end{equation}

with $p_0=4E_0\beta c^2/\pi^{3/2}C_p{d_0}^3$ and $\mathbf{R}=\mathbf{r}-\mathbf{r}'$. Here $E_0$ $\beta$, $c$, and $C_p$, are the the energy per pulse, the thermal expansion coefficient, the sound speed medium propagation and the heat capacity at constant pressure of the fluid sample, respectively. In Sigrist paper, $d_0$ was defined as the spatial illumination profile of a Gaussian beam~\cite{Sigrist:1978}. However, we associate this parameter with the thermal size of the object. This hypothesis is justified from the assumption that a NP only can absorb radiation, due to plasmonic effect~\cite{Zhong:2015}, meanwhile the surrounding fluid medium (water) does not. According to this, a minimum value for $d_0$ is the NP diameter ($d_\text{NP}$); and as maximum the quantity $(d_\text{NP}+d_{th})$ where:
\begin{equation}\label{eq02}
d_{th}=4 (\chi_w\tau_l)^{\frac{1}{2}}.
\end{equation}
Equation \eqref{eq02} is related the thermal diffusion length~\cite{Carslaw}; for this expression $\chi_w$ is the water diffusivity ($0.143\times10^{−6}\text{m}^2/\text{s}$) and $\tau_l$ is the laser pulse (FWHM of 10 ns); therefore, $d_{th}=150$ nm. This analytical approximation is equivalent to solve the coupling heat and pressure equations considering the laser time profile.

A code was written in the software Wolfram Mathematica\textsuperscript{TM} to emulate the experiments performed. Considering the equation \eqref{eq01} and the optical attenuation, the numerical expression employed was: 
\begin{equation}\label{eq03}
p_{sim}(\mathbf{D}_{ij},t)=\sum^{n, m}_{i=1,j=1}10^{- \epsilon \zeta j \Delta z}\times p(\mathbf{D}_{ij},t);
\end{equation}
with $\mathbf{D}_{ij}=\mathbf{s}-\mathbf{r}_{ij}$. Here $\epsilon$ corresponds to the extinction coefficient and $\zeta$ is the sample concentration, which were measured for each NPs aliquot by UV-VIS spectroscopy. Likewise , $\mathbf{s}$ corresponds to the sensor position. Counters $i$ and $j$ are used to label each NP and the cylinder section, respectively, the length of the sections is $\Delta z$. The complete methodology details used to perform the simulations can be found in the supplemental material.

In Figures \ref{Fig3}a to \ref{Fig3}c simulated PA signals for stock concentration and $d_0=d_\text{NP}$ are presented. A statistical study shows that all signals are symmetric with well defined maximum and minimum peaks, which always appear near the center temporal range. To obtain the P-P PA amplitude it was sought the higher and lower peaks values for each individual simulation. Same behavior was observed when the BL law is considered, nonetheless, the maximum amplitude is diminished, as expected, being 60\% less when compared to the case without attenuation.
\begin{figure}[h!]
	\centering
	\includegraphics[width=0.48\textwidth]{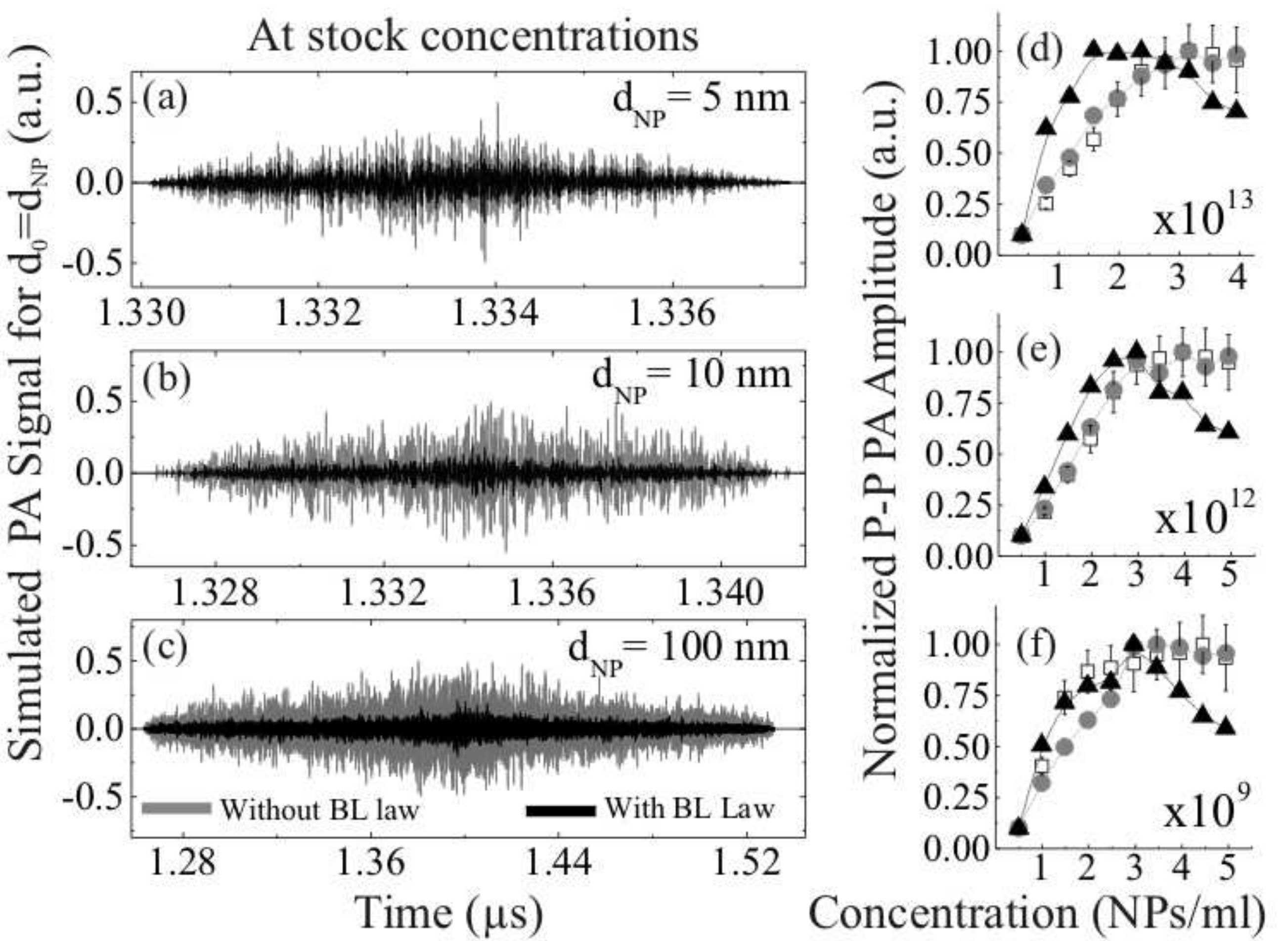}\label{Fig3}
	\caption{(a)-(c) Simulated PA signals for $d_0=d_\text{NP}$ at stock concentration for NPs of $5$ nm, $10$ nm  and $100$ nm respectively; the gray lines are the simulations without BL law and the black ones with BL. (d)-(f) Corresponding comparison between simulated and experimental P-P PA amplitudes as a function of concentration, squares for experimental data, circles for simulations without BL law and triangles for simulations with BL law. All data is normalized.}
    \label{Fig3}
\end{figure}

In Figures \ref{Fig3}d-\ref{Fig3}f, a comparison between the experimental and simulated P-P PA amplitudes as a function of the NPs concentration is shown; all data were normalized to the respective maximum amplitude. Simulations for $d_0=d_\text{NP}$ without BL law predict completely the experimental trend, but when the optical attenuation is considered it fails; contrary to the expected results, the inclusion of BL law in the model did not predict the experiments. Trying to understand this discrepancy, the numerical Fourier transform can be performed to the simulated signals, when doing this a broad spectra are predicted with high central frequencies for $d_0=d_\text{NP}$. The frequencies are at 100 GHz for the $5$ nm and $10$ nm samples and at 10 GHz for $100$ nm; however, this is not in agreement with the actual spectral response of the sensors that we used in our experiments, then the $d_0=d_{\text{NP}}$ assumption must be discarded. 

In Figures \ref{Fig4}a to \ref{Fig4}c, simulations supposing $d_0=d_\text{NP}+d_{th}$ are shown for the stock concentration of NPs. There are two aspects that must be highlighted: First, amplitude-shape for $5$ and $10$ nm samples are well-defined for all simulated concentrations; however, for $100$ nm NPs it remains noisy, but its shape is more defined than for the case $d_0=d_\text{NP}$. Second, opposite to the above case, PA signals are asymmetric for all samples, being like the experimental results and the reported literature. Considering BL law still decreases the PA amplitude approximately at 40\% of the non-attenuated value. Comparison between experimental and simulated P-P PA amplitudes as a function of the NPs concentration is presented in Figures \ref{Fig4}d to \ref{Fig4}f. The simulated PA signals without BL law are linear. When light attenuation is considered, behavior of experimental signals is well predicted; as in the previous cases, the spectra for the simulated signals can be calculated numerically, the obtained frequencies now are between $1$ and $100$ MHz. For both sets of simulations, optical attenuation, through BL law, defines the amplitude of pressure wave and its effect is only to decrease the total PA amplitude. 
\begin{figure}[h!]
	\centering
    \includegraphics[width=0.48\textwidth]{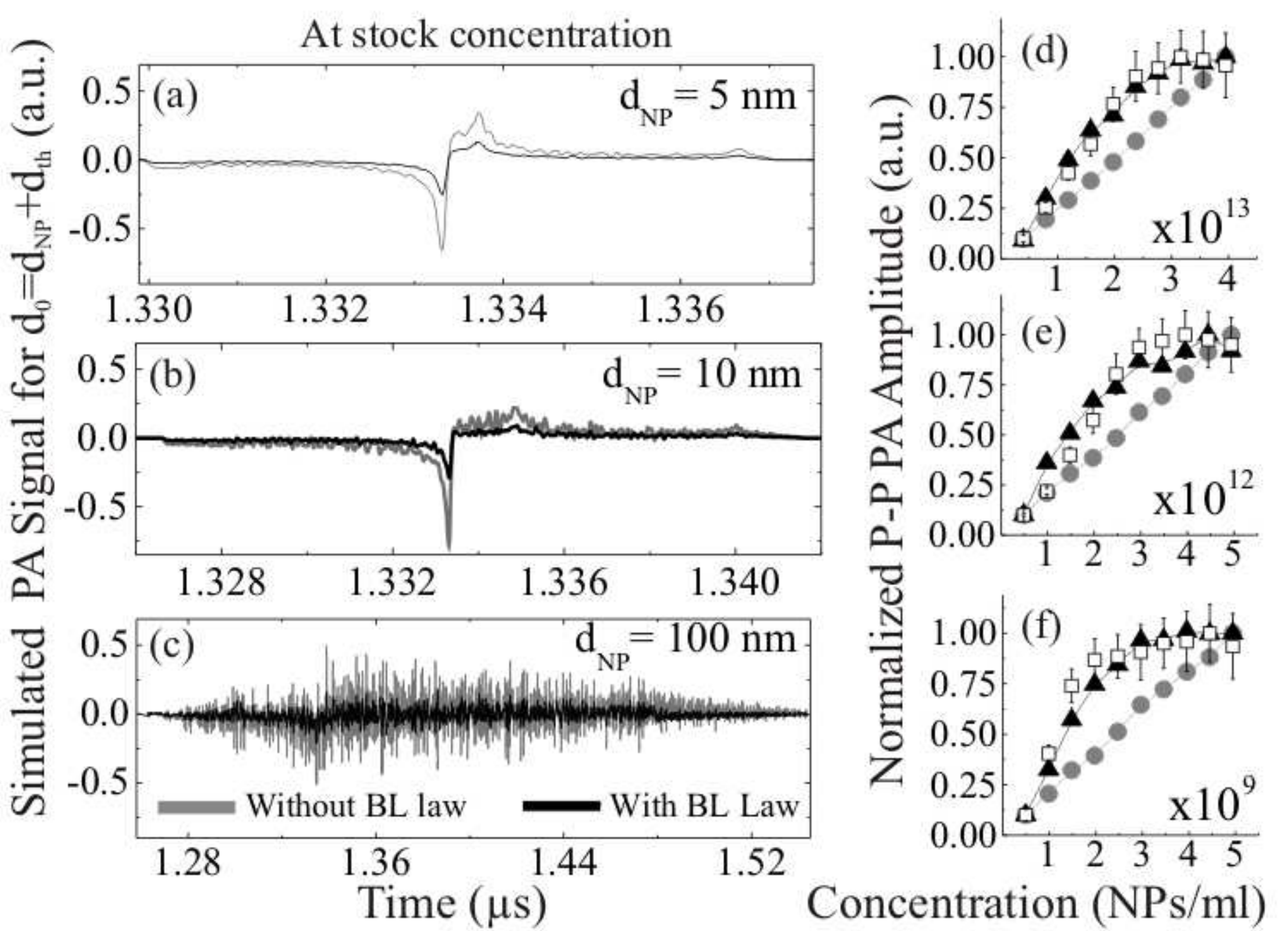}\label{Fig4}
    \caption{(a)-(c) Simulated PA signals for $d_0=d_\text{NP}+ d_{th}$ at stock concentration for NPs of $5$ nm, $10$ nm  and $100$ nm respectively; the gray lines are the simulations without BL law and the black ones with BL. (d)-(f) Corresponding comparison between simulated and experimental P-P PA amplitudes as a function of concentration, squares for experimental data, circles for simulations without BL law and triangles for simulations with BL law. All data is normalized.}
    \label{Fig4}
\end{figure}

\begin{figure}[h!]
	\centering
    \includegraphics[width=0.48\textwidth]{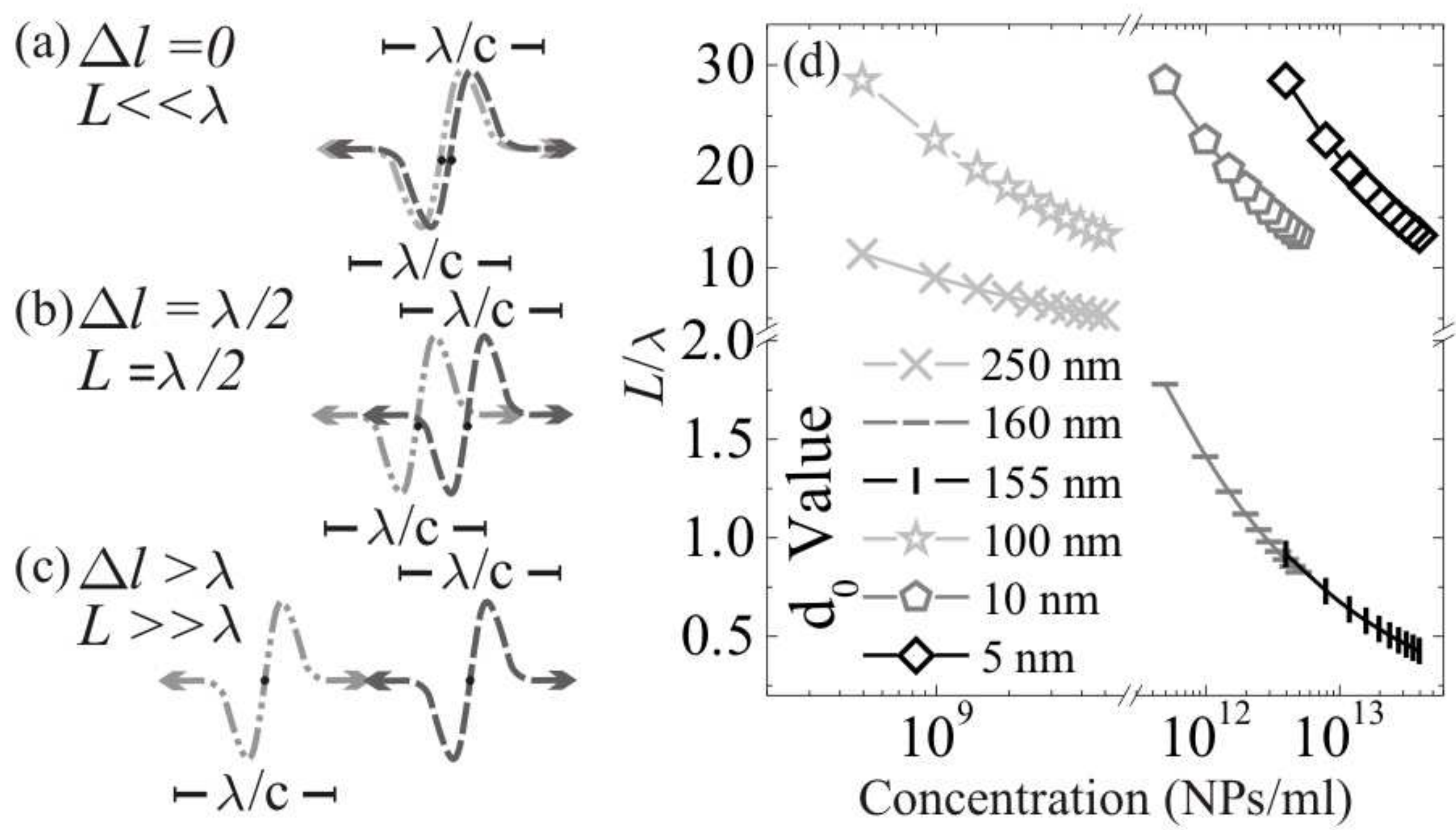}
    \caption{(a)-(c) Present the tree extreme possible configuration between the sources, totally constructive interference, partial destructive interference and when the signals can be superposed, respectively. (d) Ratios $L/\lambda$, for all $d_0$ values used in the simulations. }
    \label{Fig5}
\end{figure}

To better understand the consequences for choosing a $d_0$ value, three aspects must be considered. First, from the PA power spectrum of a single NP it is found that the maximum frequency value occurs at $\nu_{\text{max}}=c/\sqrt{2}\pi (d_0/2)$. From this value can be calculated the spatial region where pulse of one NP can interact with each other, it corresponds to $\lambda\equiv\sqrt{2}\pi d_0$.  Second, from the specific volume of the NPs suspension, an average distance between NPs $L$, can be determined. Third, equation \eqref{eq01} is proportional to the time derivative of a Gaussian function, which has a bipolar temporal profile i.e., it is compose of a compression and rarefaction cycle. When summation over two individual PA signals is performed at the measuring point, there are three extreme possible situations for the time delay (or acoustical path difference $\Delta l$), namely: (i) it is equal to zero, then the PA pulses match exactly and only constructive interference appears. (ii) it is equal to $\lambda/2c$; then, the rarefaction of one pulse corresponds exactly to the compression of other pulse, and therefore partial destructive interference is produced. (iii) It is greater than $\lambda/c$, so they cannot superimpose. These cases are displayed in Figures \ref{Fig5}a to \ref{Fig5}c. Using the above information, the ratios $L/\lambda$ where calculated for all $d_0$ values and are show in Figure \ref{Fig5}d. When $d_0=d_\text{NP}$, $L/\lambda\gg1$ for all NPs diameters and all NPs concentrations, thus a high number of NPs cannot be superimposed; therefore, the sum of the individual signals, at the measurement point during a time interval, looks noisy and symmetric. For $d_0=d_\text{NP}+d_{th}$, we can see in Figure \ref{Fig5}d that $L/\lambda<2$ for $d_\text{NP}= 5$ nm and $d_\text{NP}=10$ nm, respectively; now summation over individual signals produces well defined shape and asymmetric PA signals with linear behavior of the P-P PA amplitudes as a function of the NP’s number. This is because at the measurement point, in the time interval, the superposition of individual pressure waves occurs. For $d_0=100$ nm$+d_{th}$ the corresponding ratio is in the range $5<L/\lambda<12.5$, then the interference is more probable than the case $d_0=100$ nm, but less when $d_0=d_\text{NP} +d_{th}$ for $5$ nm and $10$ nm. The consequence for adding individuals PA pulses, considering the heat diffusion from the NP volume to their surroundings, gives a PA signal with high signal-to-noise ratio, asymmetric shape peaks and a linear behavior in the P-P PA amplitude as a function of the NP’s number. When the optical attenuation is taking into account the simulated PA amplitudes loss its linear dependence with the NPs concentration and the signal saturates, predicting properly the experimental results.

It is important to remark that, if the results are displayed as a function of the NPs OD instead of their concentrations, the nonlinear behavior of the P-P PA amplitudes is given for OD$\geq0.5$. A meticulous review of the references \cite{Huang:2013,Bayer:2013,Solano:2012,Karpiouk:2008,Saha:2011,Mallidi:2009,Chamberland:2008,Jokerst:2012,Galanzha:2009,Khosroshahi:2015,Song:2009,Pan:2010,WangJ:2010} is in concordance with this threshold (see supplementary data). 

In summary, when heat propagates beyond the individual NP’s volume and the optical attenuation of the sample is ignored, the P-P PA amplitude as a function of NP’s concentration is linear; this extended PA source can improve the interference between the single US pulses. The simulations showed that asymmetric shape of the PA signal is obtained under this condition. When the heat is confined inside the NP dimensions, symmetric signals are obtained and a nonlinear P-P PA amplitude; the NP thermal confinement can be discarded through frequency spectrum too. The saturated behavior of P-P PA amplitude for the extended thermal source is correctly explained when the optical attenuation is considered. Finally, our simulations and experimental results showed that no linear behavior appears for an OD$\geq0.5$. This threshold was well-matched with previous experimental reports. This value can be taken as a point of departure to obtain linear PA amplitudes as a function of the concentration for NPs samples.

\textbf{Acknowledgments}. The computation for this work was performed on the high-performance computing infrastructure provided  by Research Computing Support Services and in part by the National Science Foundation under grant number CNS-1429294 at the University of Missouri, Columbia Mo. The experimental part of this work was financed with the 2nd edition of UG-CIO research grants and CONACyT (grant number 5215708). Authors thank to Martin Olmos and Enrique Noe Arias for their support in this research and CONACyT.

\bibliography{bibl}

\end{document}